# Characteristics of ion-acoustic solitary wave in a laboratory dusty plasma under the influence of ion-beam

M K Deka[1], N C Adhikary[1, a)], A P Misra[2], H Bailung[1] and Y Nakamura[3]

[1] *Physical Sciences Division, Institute of Advanced Study in Science and Technology, Vigyan Path, Paschim Boragaon, Garchuk, Guwahati-781035, Assam, India.*
[2] *Department of Mathematics, Siksha Bhavana, Visva-Bharati University, Santiniketan-731 235, India.*
[3] *Department of Physics, Faculty of Engineering, Yokohama National University, 79-5 Tokiwadai Hodogaya-ku, Yokohama 240-8501, Japan.*

a) Corresponding Author's E-mail: nirab_iasst@yahoo.co.in.

**Abstract**: We study the influence of ion beam and charged dust impurity on the propagation of dust ion-acoustic (DIA) solitary wave (SW) in an unmagnetized plasma consisting of Boltzmann distributed electrons, positive ions, positive ion beam and negatively charged immobile dusts in a double plasma device. On interacting with an ion beam, the solitary wave is bifurcated into a compressive fast and a rarefactive slow beam mode, and appears along with the primary wave. However, there exists a critical velocity of the beam beyond which the amplitude of the fast solitary wave starts diminishing and rarefactive slow beam mode propagates with growing amplitude. Whereas, the presence of charged dust impurity in the plasma reduces this critical beam velocity and a substantial modification in the phase velocity of the slow beam mode is observed with increasing dust density. Furthermore, the nonlinear wave velocity (Mach number) as well as the width of the compressive solitons are measured for different beam velocity and dust density, and are compared with those obtained from the K-dV equation. The experimental results are found in a well agreement with the theoretical predictions.

## I. INTRODUCTION

Low temperature plasmas containing massive charged dusts have attracted a considerable attention over the last few decades. Such plasmas are relevant in astrophysical settings, such as formation of stars in interstellar clouds or physical processes in cometary tails[1,2] and planetary magnetospheres[3] as well as in industrial plasmas, e.g., semiconductor processing[4,5] etc.. The presence of these charged dusts in electron-ion plasmas modifies the propagation of dust ion-acoustic wave (DIAW) or dust-acoustic wave (DAW) depending on whether the charged dusts are static or mobile. Since the first theoretical prediction of such waves by Shukla and Silin[6] followed by experimental observations by Barkan *et al.*[7], a number of works have been done by several authors both theoretically[8-12] and experimentally[13,14] to study various features of DIA and DA waves in different plasmas. Investigations on DIAWs in nonplanner geometry (cylindrical and spherical) using the reductive perturbation method reveal that their characteristics differ from those in a planner one dimensional geometry [15,16]. Mamun and Shukla[17] also reported that in such a nonplanner geometry, nonthermal electron distribution and polarity of net charge number density of dust particles can appreciably alter the DIAWs structure.

On the other hand, energetic charged particles such as ion beam can significantly modify the propagation characteristics of solitary waves in plasmas[18, 19]. Solitary waves with negative potential have been found in the vicinity of ion beam regions of the auroral zone in the upper atmosphere[20]. The spacecraft observation in the Earth's plasma sheet boundary indicates that the electrons and ion beams can also drive the broadband electrostatic waves[21]. Also, it has been observed that the propagation of high current finite length ion beam in the background plasma plays a significant role in many applications, e.g., heavy-ion fusion, plasma lenses, cosmic ray propagation etc.[22]. The stabilization of linear ion beams in space plasmas or launched wave packets in ion beam plasmas are problems of great importance[23].

It is well known that when an ion beam is injected into an unmagnetized plasma involving the ion motion, three longitudinal electrostatic modes, namely an ion-acoustic wave and a fast and slow space charge waves, can propagate in the plasma. The nonlinear characteristics of these three modes were theoretically investigated by Yajima *et al.*[24]. They predicted that each mode has soliton solutions of the Korteweg–de Vries (K-dV) type under the assumption that the beam density is much smaller than the plasma density and that the wave amplitude is much smaller than $\kappa_B T_e / e$, where $\kappa_B$ is the Boltzmann constant, $T_e$ is the electron temperature and $e$ is the elementary charge. In their investigation they had neglected the temperatures of both the plasma and the beam ions, and considered Boltzmann distribution for electrons. The properties of K-dV type solitons in beam plasmas have been investigated by Gell *et al.*[25] and El-Labany[26] taking into account the effects of finite temperatures of all the species.

The effects of both the background and beam ion temperatures in such a plasma were later analyzed by Nakamura and Ohtani [27]. They concluded that compressive (positive) solitary waves can exist for these three modes when they are stable. But, no solitary wave propagates in the unstable case. The later is caused by the ion-ion instability in which ion acoustic wave (IAW) couples with the slow beam mode, and the instability occurs when the beam velocity is (1~2) times the ion-acoustic velocity, and the beam density is much smaller than the plasma density. This ion-ion instability has been confirmed experimentally by Gresillon and Deveil [28]. These results were also in good agreement with those obtained by Yajima *et al.*[24]. Nevertheless, Ohnuma and Hatta[29] first investigated the instability of these wave modes in unmagnetized beam plasmas, mainly by means of graphical analysis of the dispersion relation. They found that the ratio $\sigma_b = T_b/T_e$ played an important role in determining the unstable region of ion modes, where $T_b$ is the beam ion temperature.

The role of adiabaticity of positive ions and electrons on the propagation of small amplitude DIAWs with charged dust impurities have been discussed by some researchers and it has been reported that the combined effects of adiabatic electrons and negatively charged static dust significantly modify the propagation characteristics (speed, amplitude, and width) of the K-dV DIA solitons[30,31]. However the effects of adiabatic positive ions and beam ions as well as their finite temperatures on the propagation of small amplitude DIAWs with charged dust impurities have not yet been considered and discussed in detail. In this work, a theoretical as well as an experimental study of the propagation of DIAWs in an ion beam driven dusty plasma carried out in a dusty double plasma device is presented. The observed results are found to be in good agreement with the theoretical predictions. The present manuscript is organized as follows; in section II detail theoretical description is provided, section III contains the description of the experimental set up, in section IV results and discussion are presented and finally conclusion is given in the section V.

## II. THEORETICAL FORMULATION

We consider an unmagnetized plasma composed of positive ions, positive ion beam, Boltzmann distributed electrons, and negatively charged massive dust grains forming only the background plasma. Here two distributions for ions; namely the bulk uniform warm ions with no equilibrium flow and the energetic ion beam with an equilibrium velocity $V_{b0}$ has been considered. The model thus advances the previous work of Adhikary *et al.* [17] with the fact that the finite temperature of the plasma ions and beam ions are taken into account. The normalized set of basic equations describing the propagation of DIA waves in such plasma can be expressed as

$$\frac{\partial N_i}{\partial T} + \frac{\partial}{\partial X}(N_i V_i) = 0 \qquad \text{---------------------- (1)}$$

$$\frac{\partial V_i}{\partial T} + V_i \frac{\partial V_i}{\partial X} + \frac{\sigma_i}{N_i}\frac{\partial N_i}{\partial X} = -\frac{\partial \Phi}{\partial X} \qquad \text{---------------------- (2)}$$

$$\frac{\partial N_b}{\partial T} + \frac{\partial}{\partial X}\left(N_b V_b\right) = 0 \qquad \text{---------------------- (3)}$$

$$\frac{\partial V_b}{\partial T} + V_b \frac{\partial V_b}{\partial X} + \frac{\sigma_b}{\mu_m N_b}\frac{\partial N_b}{\partial X} = -\mu_m \frac{\partial \Phi}{\partial X} \qquad \text{---------------------- (4)}$$

$$\frac{\partial^2 \Phi}{\partial X^2} = e^{\Phi} + \mu_d - N_b \mu_b - N_i \mu_i \qquad \text{---------------------- (5)}$$

where $N_i$ ($N_b$) denotes the number density of positive ions (beam ions) normalized by their equilibrium values $n_{i0}$ ($n_{b0}$), $V_i$ ($V_b$) denotes the velocity of positive ions (beam ions) normalized by the ion sound velocity $C_s = (\kappa_B T_e / m_i)$, $\Phi$ is the dust ion acoustic wave potential normalized by $\kappa_B T_e / e$, and $\sigma_i = T_i / T_e$ is the ion to electron temperature ratio. Furthermore, $\mu_m = m_i / m_b$ is the plasma ion to beam ion mass ratio, $\mu_d = Z_d n_{d0} / n_{e0}$ is the ratio of the equilibrium dust density (multiplied by $Z_d$, the number of electrons residing on the dust grains) to plasma electron density, $\mu_b = n_{b0} / n_{e0}$ is the ion beam to plasma ion density ratio and $\mu_i = n_{i0} / n_{e0}$ is the positive ion to electron density ratio. The space ($x$) and time (t) variables are respectively normalized by the Debye length, $\lambda_d = (\kappa_B T_e / 4\pi n_i e^2)^{1/2}$ and the inverse of the plasma frequency $\omega_{pi} = (4\pi n_i e^2 / m_i)^{1/2}$. Therefore, at equilibrium, the overall charge neutrality condition is

$$1 - \mu_i - \mu_b + \mu_d = 0 \qquad \text{---------------------- (6)}$$

In order to derive the evolution equation for the propagation of small but finite amplitude DIASW, we use the standard reductive perturbation technique in which the independent variables are stretched as $\xi = \varepsilon^{1/2}(x - v_0 t)$ and $\tau = \varepsilon^{3/2} t$. The dependent variables $N_i$, $V_i$, $N_b$, $V_b$ and $\Phi$ can be expanded in power series of $\varepsilon$ as

$$N_i = 1 + \varepsilon N_i^{(1)} + \varepsilon^2 N_i^{(2)} + \ldots\ldots\ldots\ldots \qquad \text{---------------------- (7)}$$

$$V_i = 0 + \varepsilon V_i^{(1)} + \varepsilon^2 V_i^{(2)} + \ldots\ldots\ldots\ldots \qquad \text{---------------------- (8)}$$

$$N_b = 1 + \varepsilon N_b^{(1)} + \varepsilon^2 N_b^{(2)} + \ldots\ldots\ldots\ldots \qquad \text{---------------------- (9)}$$

$$V_b = V_{b0} + \varepsilon V_b^{(1)} + \varepsilon^2 V_b^{(2)} + \ldots\ldots\ldots\ldots \qquad \text{---------------------- (10)}$$

$$\Phi = 0 + \varepsilon \Phi^{(1)} + \varepsilon^2 \Phi^{(2)} + \ldots\ldots\ldots\ \ldots\ldots\ldots\ . \qquad \text{---------------------- (11)}$$

where $\varepsilon$ is a small nonzero constant measuring the weakness of the dispersion and $v_0$ is the nonlinear wave velocity (Mach number) normalized by the ion-sound velocity $C_s$.

Substituting the stretched co-ordinates and the expressions for $N_\alpha$ and $V_\alpha$ and $\Phi$ into the basic equations (1) to (5) and equating the coefficients of lowest order of $\varepsilon$ we get,

$$N_i^{(1)} = \Phi^{(1)} / \alpha = V_i^{(1)} / v_0 \qquad \text{-------------------- (12)}$$

$$N_b^{(1)} = \mu_m \Phi^{(1)} / \beta = V_b^{(1)} / (v_0 - V_{b0}) \qquad \text{-------------------- (13)}$$

$$\Phi^{(1)} = \mu_i N_i^{(1)} + \mu_b N_b^{(1)} \qquad \text{-------------------- (14)}$$

together with the dispersion relation

$$v_0^2 = \frac{1}{2(1-\gamma)^2}\left[\left\{\frac{\sigma_b}{\mu_m} + \sigma_i(1-\gamma)^2 + \mu_i(1-\gamma)^2 + \mu_b\mu_m\right\}\right.$$

$$\left.\pm\sqrt{\left\{\frac{\sigma_b}{\mu_m} + (\sigma_i+\mu_i)(1-\gamma)^2 + \mu_b\mu_m\right\}^2 - 4(1-\gamma)^2\left\{\frac{\sigma_i\sigma_b}{\mu_m} + \mu_i\frac{\sigma_b}{\mu_m} + \mu_b\mu_m\sigma_i\right\}}\,\right] \qquad \text{----- (15)}$$

where, $\alpha = (v_0^2 - \sigma_i)$, $\beta = \{(v_0 - V_{b0})^2 - (\sigma_b/\mu_m)\}$

and $\gamma = 1 - (V_{b0}/v_0)$

Proceeding this way we finally we get the following K-dV equation

$$\frac{\partial\Phi}{\partial\tau} + A\Phi\frac{\partial\Phi}{\partial\xi} + B\frac{\partial^3\Phi}{\partial\xi^3} = 0 \qquad \text{-------------------- (16)}$$

where $\Phi \equiv \Phi^{(1)}$ and the nonlinear co-efficient A and dispersive co-efficient B are given by

$$A = B\left[-1 + \frac{\mu_i(3v_0^2 - \sigma_i)}{(v_0^2-\sigma_i)^3} + \frac{\mu_b\left[(2+\mu_m)(v_0-V_{b0})^2 - \mu_m\sigma_b\right]}{\{(v_0-V_{b0})^2 - \sigma_b\}^3}\right] \qquad \text{-------------------- (17)}$$

$$B = \left[\left(\frac{(v_0^2-\sigma_i)\{(v_0-V_{b0})^2-\sigma_b\}^2}{2\mu_i v_0\{(v_0-V_{b0})^2-\sigma_b\}^2 + (v_0^2-\sigma_i)\{2\mu_b(v_0-V_{b0})\}}\right)\right] \qquad \text{-------------------- (18)}$$

The stationary soliton solution of the K-dV equation is obtained by transforming independent variables $\xi$ and $\tau$ to a single new variable $\zeta = \xi - U_0\tau$, where $U_0$ is the constant phase velocity (normalized by $C_s$), and imposing the appropriate boundary conditions for localized perturbations (viz., $\Phi \to 0$, $\partial\Phi/\partial\xi \to 0$ and $\partial^2\Phi/\partial\xi^2 \to 0$ and as $\xi \to \pm\infty$) as $\Phi = \Phi_m \sec h^2[(\xi - U_0\tau)/D]$. The amplitude $\Phi_m$ (normalized by $\kappa_B T_e/e$) and the width *D* (normalized by $\lambda_d$) of the soliton are given by $\Phi_m = 3U_0/A$ and $D = \sqrt{4B/U_0}$.

We note that for the soliton solutions to exist the coefficient *B* must be positive. Both *A* and *B* typically depend on the temperature ratios, the beam velocity as well the ratios of the number densities including the contribution from the charged dusts. The behaviors of *A* and *B* as dependent on $\mu_d$ are analyzed numerically for different values of the beam velocity, and are shown in Figures 1 (a) & 1 (b). It is found that the nonlinearity effect is enhanced with the beam velocity, whereas it goes down with the increase in the charged dust concentration. On the other hand, the dispersion coefficient *B* increases with $\mu_d$, but decreases with increasing the beam ions velocity.

## III. EXPERIMENTAL SET UP AND MEASUREMENT

The experiment is performed in a dusty double-plasma device made of a stainless steel cylindrical chamber having 120 cm length and 30 cm in diameter. This chamber is equipped with multi-dipole permanent magnets cage for surface confinement [15] as shown in Figure 2. The device is separated into source and target section by a mesh grid which is kept floating. Dusty plasma is produced by sufficient amount of dust into the plasma by means of a dust dispersing setup. This dust dispersing setup is fitted at the top side of the target section of the chamber consists of an ultrasonic vibrator coupled with a dust reservoir; and the amount of dispersed dust can be controlled by the applied voltage into the ultrasonic vibrator.

The chamber is evacuated down to the pressure of 1.6 $\times 10^{-6}$ Torr with a diffusion pump backed by a rotary pump. Xenon (Xe) gas is fed through a needle valve into the chamber and the pressure is maintained at $2\times 10^{-4}$ Torr under continuous pumping. In the source and target section of the chamber the Xenon plasma is produced by dc discharge between tungsten filaments of 0.1 mm diameter as cathode and the magnetic cages as anode. The discharge voltage and current are kept fixed at 60 V and 50 mA respectively. Plasma parameters are measured with the help of a plane Langmuir probe of 6 mm in diameter and a retarding potential energy analyzer (RPA) of 20 mm diameter. The RPA, which consists of two mesh grids and a collector plate, is used to estimate the ion temperature ($T_i$), ion beam energy and the densities. The position of the Langmuir probe is changeable axially by a motor driving system so that the received signal can be observe at any position. The typical plasma parameters in this experimental condition are: the electron density $n_e$ ~ (1.5 to 2) $\times 10^8$ cm$^{-3}$, the electron temperature $T_e$ ~ (1 to 1.5) eV, and the ion temperature $T_i$ ~ 0.1 eV. Dust particles used in this experiment are glass beads of average diameter 8.8 μm and the dust number density inside the plasma system can be varied from $1.6\times 10^3$ cm$^{-3}$ to 5 $\times 10^4$ cm$^{-3}$. Now, as soon as the dust grains are introduced into the plasma, they get charged due to the collection of electrons and ions from the bulk plasma. Hence, the electron and ion density in the plasma will be reduced. From the reduced amount of ion and electron density dust charge can be calculated [15].

The ion-acoustic perturbation is excited by applying a tone burst signal to the source anode from a function generator (Tektronix AFG 3021B) through a dc blocking capacitor. The density perturbation as a fluctuation in the electron saturation current is detected in the target section of the plasma using the plane Langmuir probe biased positively with respect to the plasma potential. Received signals are recorded by a digital storage oscilloscope (Tektronix TDS 2014B). A uniform stream of Xenon ion beam is allowed to inject from source to the target plasma section through the separation grid by applying a positive dc bias voltage to the source anode with respect to the grounded target anode. The energy of the ion beam can be varied by controlling the applied bias voltage to the source anode.

## IV. EXPERIMENTAL RESULTS AND DISCUSSIONS

In the Xenon plasma at first, the compressive solitons are excited by applying a tone burst signal of frequency ~ 30 KHz in the absence of ion beam and dust. This perturbation propagates towards the target section from grid as a fluctuation in the plasma density. This perturbation is recorded by a positively biased Langmuir probe. Figure 3 shows the temporal evolution of a compressive pulse with excitation amplitude $V_{ex}$ = 2 V detected by the Langmuir probe. Near the grid at a distance, $X$ = 2 cm the edge of the wave steepens due to nonlinearity. When the wave propagates away from the grid the role of dispersion becomes dominant and first compressive soliton appears when the nonlinearity balances with the dispersive effect ($X$ = 4 cm). At the same time a second peak begins to emerge. As the wave travels further away from the grid ($X$ > 4 cm), the pulse beaks into a train of solitons due to dispersion. The small hump seen in front of the pick may be due to reflected ions from the foot of the solitary wave front. This observation is also performed for different excitation voltages and as depicted in Figure 4, it is found that amplitude of the soliton increases with a higher excitation voltage.

Next, the influence of a positive ion beam on compressive soliton is observed by introducing a beam of $Xe^+$ ions along the direction of wave propagation. Keeping the beam to ion density ratio about 0.15, applied normalized beam velocity ($V_b/C_s = \mu$) is increased successively from 0 to 3 with small steps. In experimental representation we have considered $\mu = V_{b0} = V_b/c_s$ , where $V_b$ is experimental beam velocity, $c_s$ is experimentally calculated ion acoustic velocity from the linear propagation characteristics. In our case, initially beam velocity is very small and hence $V_{b0}$ can be considered to be nearly equal to zero. The observed signals for $V_{ex}$ = 2 V at $X$ = 6 cm under the influence of different ion beam velocity are shown in Figure 5 and is found that the amplitude of the solitary wave increased as $\mu$ is increased from 0 to 1.5. Simultaneously, the velocity of the solitary wave increases with the increase in beam velocity. However, this increasing trend in the amplitude limits up to the normalized beam velocity 1.5, beyond which the train of solitary structures starts diminishing and at $\mu$ ~1.9, wave structure is divided into two distinct parts: a fast beam mode and a slow beam mode. However, the ion-acoustic mode is rarely observed throughout the experiment except close to the exciter[32]. For a further increase in the normalized beam velocity, i.e. in the range $1.9 \leq \mu \leq 3$, the amplitude of fast beam mode is remarkably reduced compared to the slow beam mode. In the Figure 6 detected signals for an applied pulse of $V_{ex}$ = 2 V at different distances from the grid in presence of a beam with constant normalized velocity 0.84 is shown. The observed velocity of a compressive soliton in this case is higher than that in the absence of ion beam. The normalized amplitude of the soliton at $X$ = 4 cm is $|\delta n_e/n_e| \sim$ 0.23, and as the wave propagates away from grid ($X$ > 4 cm) the amplitude of pulse decreases, *i.e.*, $|\delta n_e/n_e| \sim$ 0.12 at $X$ = 8 cm. This is due to the damping and beam density depletion at large distances.

The important observation in this experiment is the effect of charged dust grains on the propagation of ion-acoustic wave in ion-beam plasmas. We observe that for a fixed dust density the solitary wave propagates at different normalized beam velocitys. The detected signals at a distance X = 6 cm from the grid for different beam velocity and in presence of dusts ($n_d = 2.34 \times 10^4$ cm$^{-3}$) is shown in Figure 7. It is found that even a small amount of dust can have an imperative role in the manifestation of slow beam mode than that is observed without dust. As described in Figure 6, that in absence of dusts the slow beam mode appears when $\mu$ ~1.9. However, in presence of a small amount of dusts ($n_d \sim 0.64 \times 10^4$ cm$^{-3}$) the slow beam mode appears even at $\mu$ ~ 1.8 and that value is further reduced to 1.6 with increasing dust density up to 2.34 ×10$^4$ cm$^{-3}$ as shown in Figure 7. And drastic modifications in the phase velocities of both the slow and fast beam modes are also observed with increasing dust density. The experiment is performed for a set of dust density with $n_d$ varying from 0.64 × 10$^4$ cm$^{-3}$ to 3.98 ×10$^4$ cm$^{-3}$ and same trend of results is observed. This can be explained from the fact that the presence of dust particles in plasma reduces both electron and ion densities, and increases the average electron temperature[33]. Therefore the corresponding ion acoustic velocity in the system increases. So, the introduction of dust particles execute a strong modification in the ion acoustic velocity with which the wave propagates under the influence of ion beam, which in turn helps in the manifestation of the slow beam mode [34].

After that the Mach velocity of the solitary wave is obtained from the relation $M = 1 + (A/3)(\delta n_e/n_e)$ where $A$ is the coefficient of nonlinearity in equation (16). In the experiment, Mach velocity ($M$) of the compressive solitons (first peak) is measured from the time of flight method and is normalized by the velocity of ion-acoustic waves. In absence of dust, the measured Mach velocities of the compressive pulses for different wave amplitudes with $\mu$ ranging from 0.84 to 1.5 are plotted in Figure 8 (a). The black (solid), red (dash) and green (dash dot) lines represent the theoretical curves of the Mach velocity and square, circle and up-triangle points represent the corresponding experimental points for $\mu$ = 0.84, 1.2 and 1.5 respectively. It is observed that in absence of dusts, the Mach velocity of the compressive solitons increases with increase in the wave amplitude for a fixed value of $\mu$. These features well agreed with the theoretical results within 10% of errors. It is also observed that with an increase in $\mu$, a corresponding increase of the Mach velocity is observed for fixed wave amplitude. As shown in Figure 1 (a) the nonlinear coefficient $A$ increases as the beam velocity increases, and as a result Mach velocity of the solitons also increases.

The comparison of measured Mach velocity in dusty plasma for different dust density ($n_d$ ~ 0.64 - 2.34 × $10^4$ $cm^{-3}$) with no dust condition, for an initial fixed value of $\mu$ (= 0.84) is shown in Figure 8 (b). The black (solid) line represents for no dust condition while the red (dash), green (dot) and the blue (dash-dot) lines represent the theoretical curves of the Mach velocity for the said range of dust density, whereas the square, circle, up-triangle and down-triangle points represent the corresponding experimental points. It is observed that the Mach velocity decreases with the enhancement of dust density in plasma even under the influence of a fixed ion beam. This is due to the fact that the increase in dust density decreases the nonlinear coefficient in K-dV equation, and hence the Mach velocity. Figure 8 (c) represents the measured Mach velocity for different values of $\mu$ (= 0.75 to 1.38) for a fixed dust density ($n_d$ = 2.34 × $10^4$ $cm^{-3}$). The black (solid), red (dash) and green (dash-dot) lines represent the theoretical curves of the Mach velocity and square, circle and up-triangle points represent the corresponding experimental points. Thus Mach velocity increases in presence of different beam velocity for a fixed dust density. On the other hand it is also seen that in absence of dust particles the rate of increase of Mach velocity with respect to $\mu$ is faster than in presence of dust particles.

Next, we experimentally measure the spatial width of the soliton through multiplying the temporal width (at half maxima) by its velocity, and then it is normalized by the Debye length ($\lambda_d$). The normalized widths $D$ of the first compressive peak measured as a function of wave amplitude for the range of $\mu$ (i.e. from 0.84 to 1.5) is shown in the Figure 9 (a). Here the theoretical curves representing the values of the width ($D$) are obtained from the solution of the K-dV equation for different conditions. The black (solid), red (dash) and green (dash-dot) lines represent the theoretical curves of $D$ and square, circle and up-triangle points represent the corresponding experimental points for the said range of $\mu$. It is observed that in absence of dust, the width of the compressive solitons decreases with the increase of the wave amplitude as well as $\mu$. This feature also agrees with the theoretical results. It is evident from the Figure 1 (b) that as the beam velocity increases the dispersion coefficient $B$ decreases, and so the width of the solitons.

The comparison of the measured width of the first peak of the solitary wave in dusty plasma for different dust density ($n_d$ ~ 0.64 - 2.34 × $10^4$ $cm^{-3}$) with no dust condition for an initial fixed value of $\mu$ (= 0.84) is shown in Figure 9 (b). The black (solid) line represents for no dust condition while the red (dash), green (dot) and blue (dash-dot) lines represent the theoretical curves, on the other hand square, circle, up-triangle and down-triangle points represent the corresponding experimental points for the said ranges of dust density and $\mu$. We find that the width of the solitary wave increases when the dust is introduced (since the phase velocity of the solitary wave increases with increasing the dust density). However, as shown in Figure 9 (c) the width of the solitary wave decreases with an increase in $\mu$ (= 0.75 to 1.38) for a fixed dust density ($n_d$ = 2.34 × $10^4$ $cm^{-3}$). The black (solid), red (dash) and green (dash-dot) lines represent the theoretical curves while square, circle and up-triangle points represent the corresponding experimental points. It is also evident that in absence of dust particles the rate of decrease of the width with respect to the ion-beam velocity is faster than in presence of dusts.

## V. CONCLUSION

We have investigated the nonlinear propagation of DIAWs in ion-beam plasmas with charged dust impurities. It is observed that the system mainly supports the propagation of compressive solitary waves. The effects of ion beam as well as the stationary charged dust grains on the propagation characteristics of compressive ion-acoustic solitary waves are studied. Due to nonlinear interaction with the ion beam, the solitary wave front appears as composed of compressive fast and a rarefactive slow beam mode along with the primary wave. A critical velocity of the beam is found to exist beyond which the amplitude of the fast beam mode tends to weaken and rarefactive slow beam mode propagates with growing amplitude. It is also observed that charged dusts can play a vital role in beam plasmas with the modification of slow as well as the fast beam modes. The velocity and width of the solitary waves are measured as a function of the wave amplitude from the temporal evolution of the wave using the Langmuir probe, and are compared with the numerical results of the K-dV equation. The experimental results show a good agreement with the theoretical estimations. The presence of dust introduces damping of the ion-acoustic waves and thereby modifies the balance between dispersion and nonlinearity. The results could be useful for the observation of solitons in space plasmas.

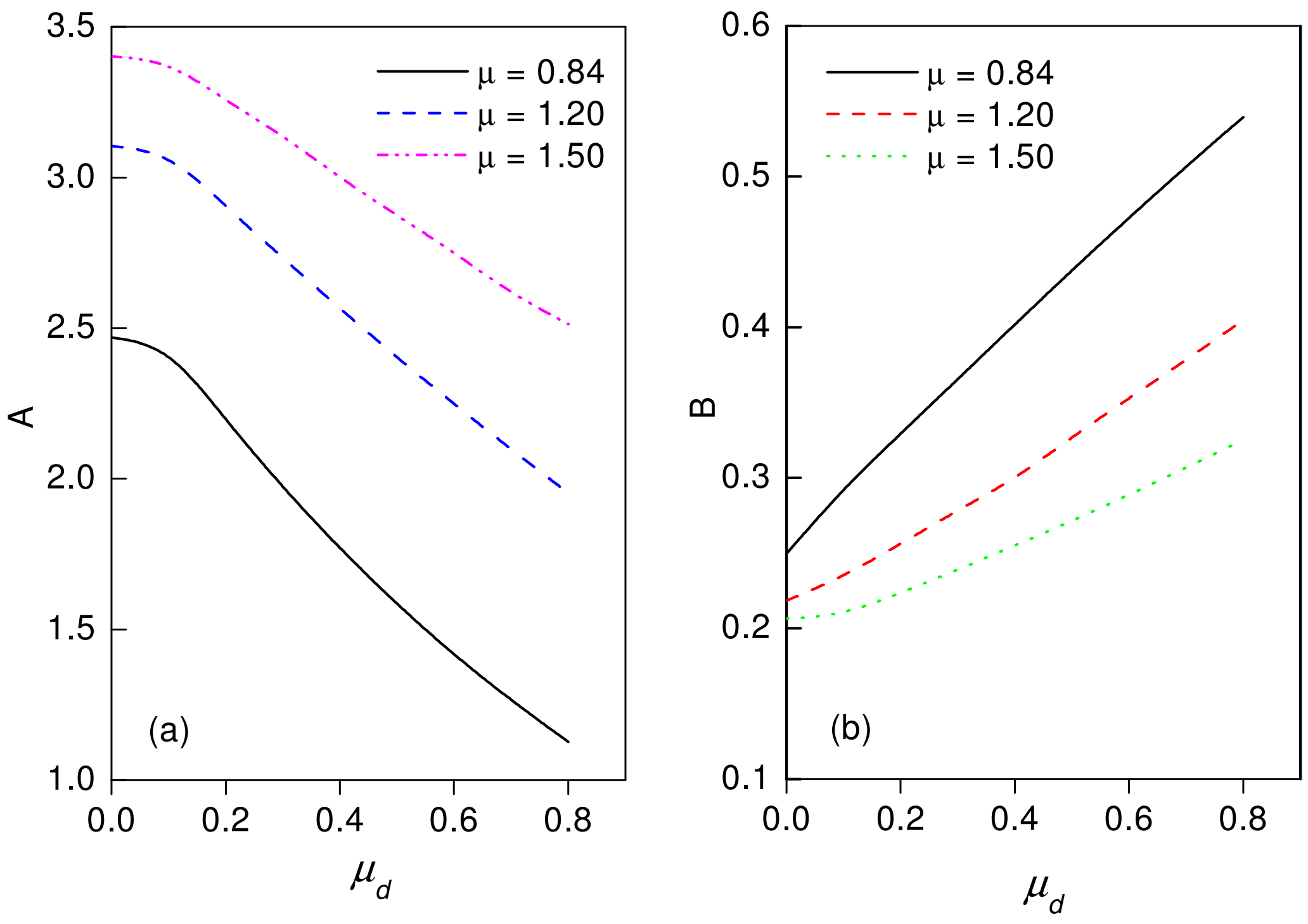

FIG. 1(a). (Color online) Variation of the nonlinear co-efficient A with $\mu_d$ for different values of $\mu$; (b) (Color online) Variation of the dispersive coefficient B with $\mu_d$ for different values of $\mu$.

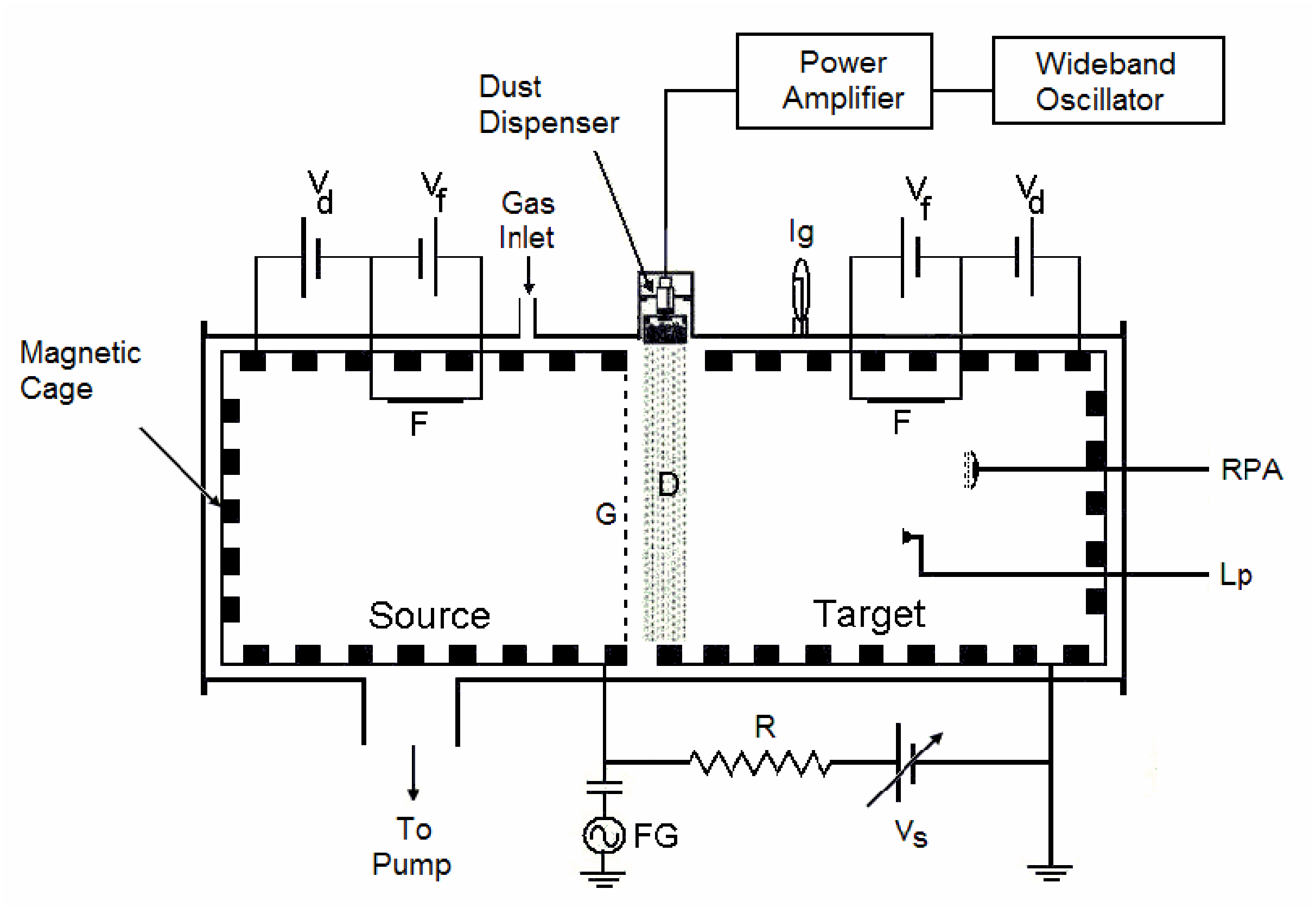

FIG. 2. Schematic diagram of the double plasma device; F - filament, G - grid, $V_f$ - filament heating power supply, $V_d$ - discharge power supply, $V_s$ - source bias, LP - Langmuir probe, RPA – retarding potential analyzer, FG- function generator, D- dust column and IG-Ionization Gauge.

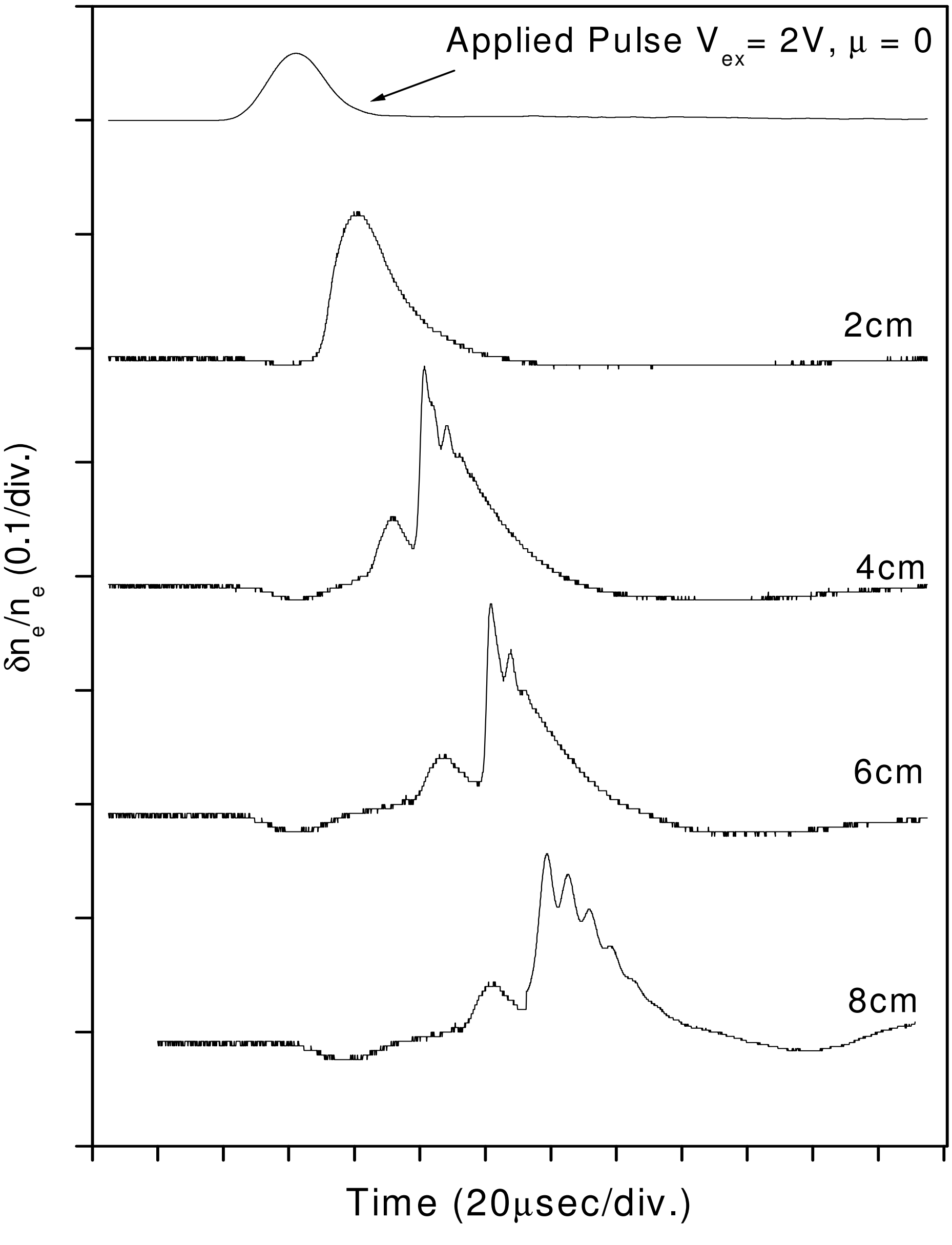

FIG. 3. Observed signals at several distances from the grid for an initial positive pulse $V_{ex}$ = 2 V in absence of ion beam. Top trace represents the applied pulse.

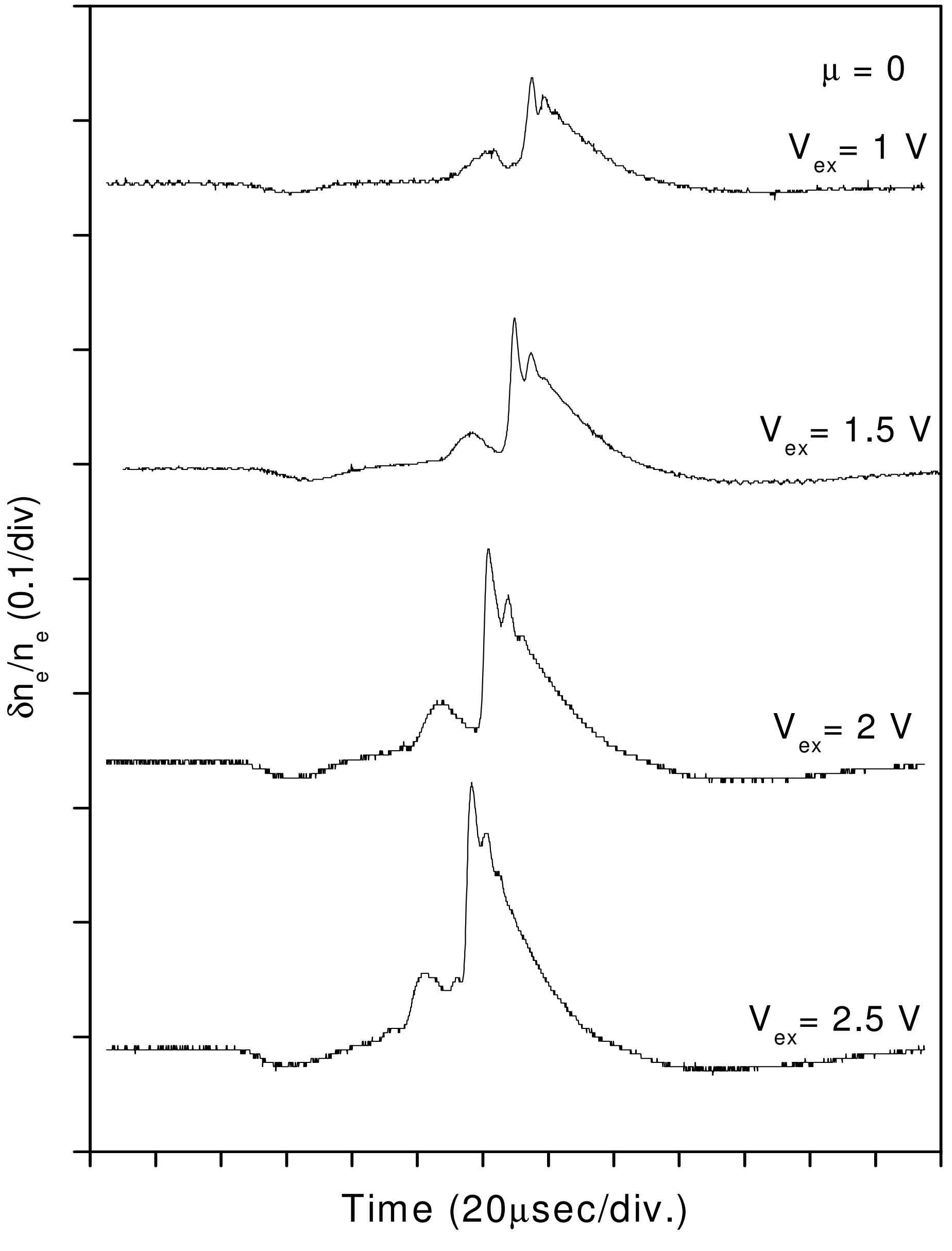

FIG. 4. Detected signals at $X = 6$ cm from the grid for different amplitudes of the positive applied pulse ($V_{ex}$) in absence of ion beam.

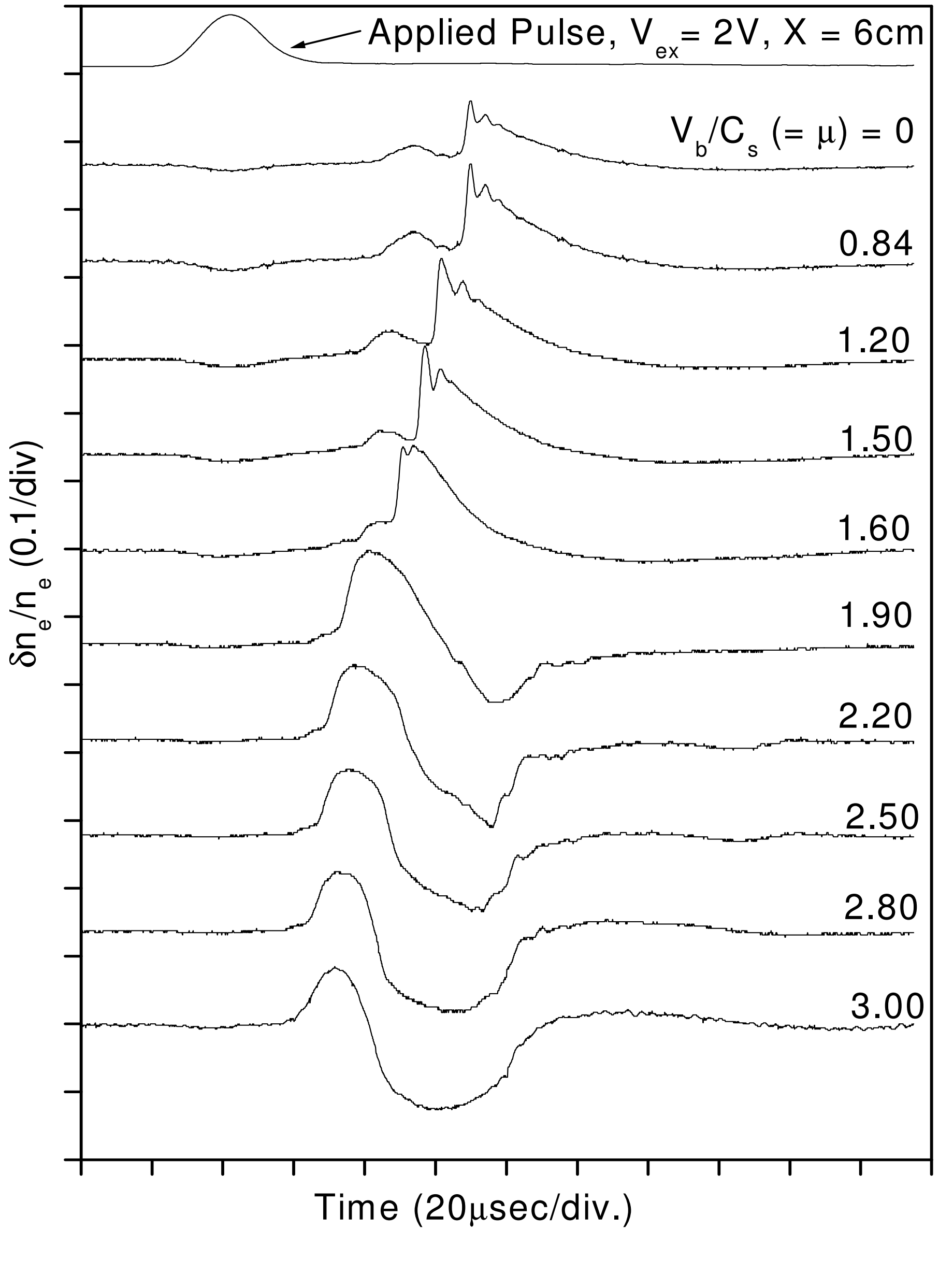

FIG. 5. Observed perturbations detected at $X = 6$ cm from the grid for an initial compressive pulse, $V_{ex} = 2$ V for different value of $\mu$. Top trace represents the applied pulse.

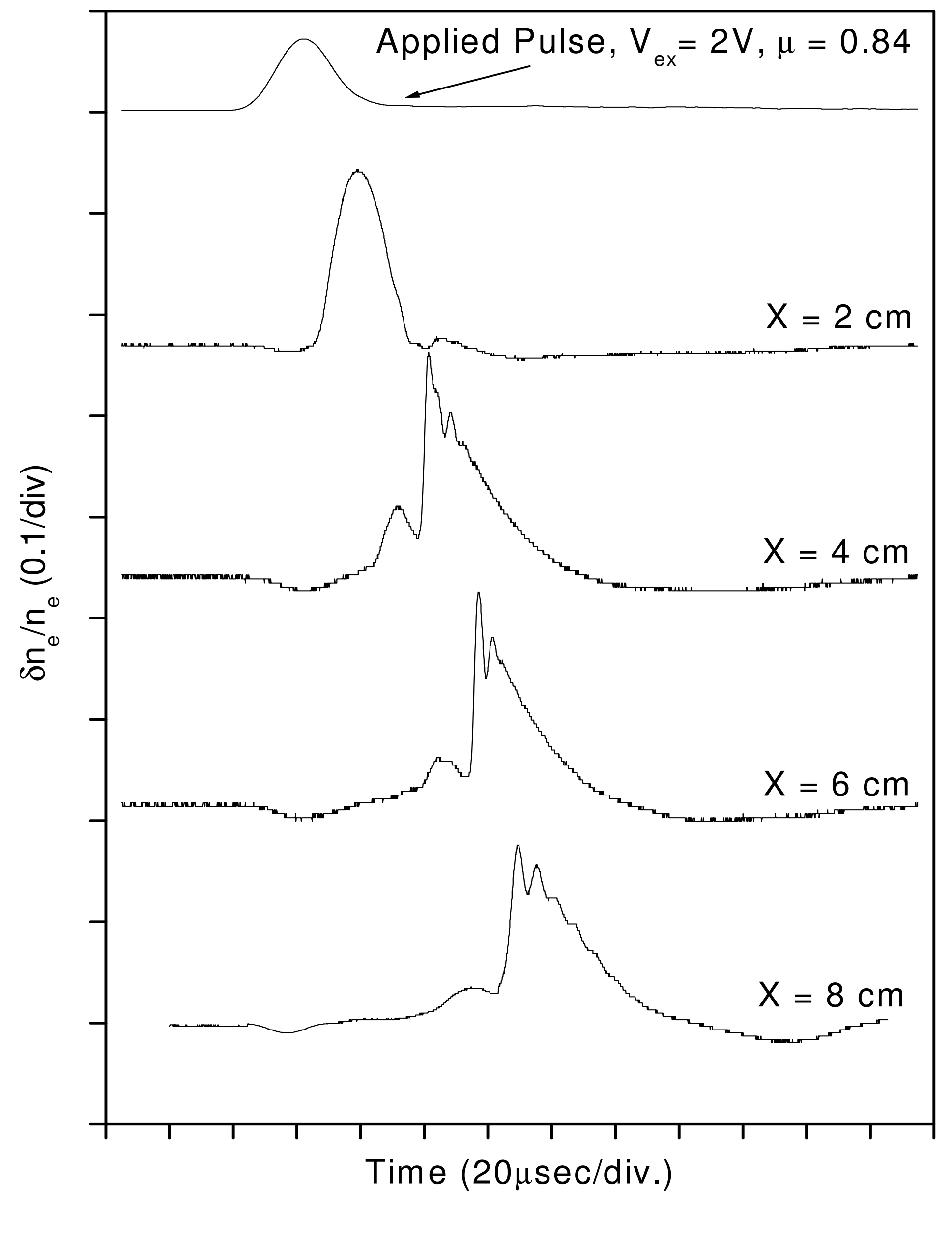

FIG. 6. Observed signals at several distances from the grid for an initial compressive pulse $V_{ex}$ = 2 V in presence of ion beam. The value of $\mu$ is fixed at 0.84.Top trace represents the applied pulse

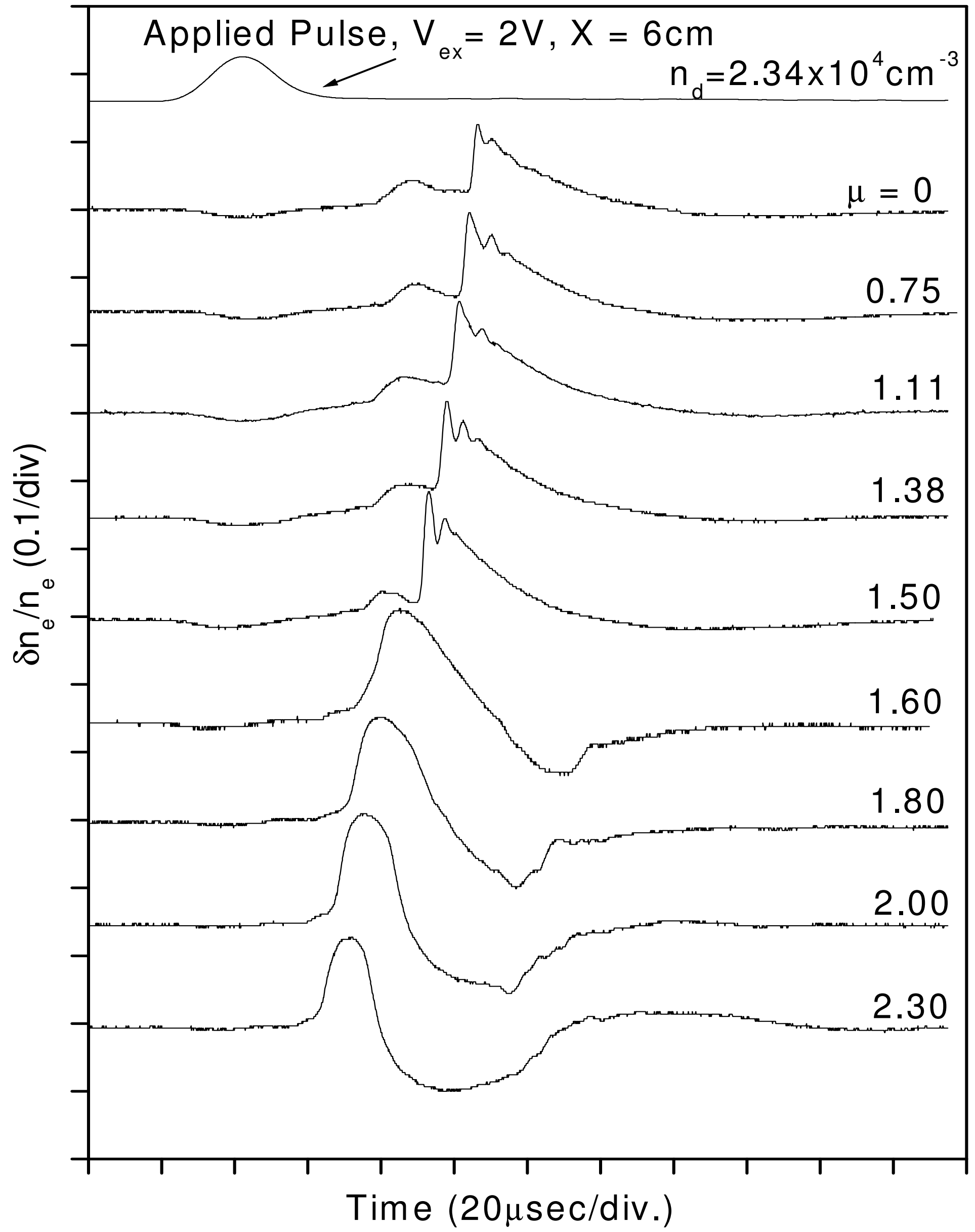

FIG. 7. Observed perturbations detected at $X$ = 6 cm from the grid for an initial compressive pulse, $V_{ex}$ = 2 V for different $\mu$ in presence of dust. Top trace represents the applied pulse.

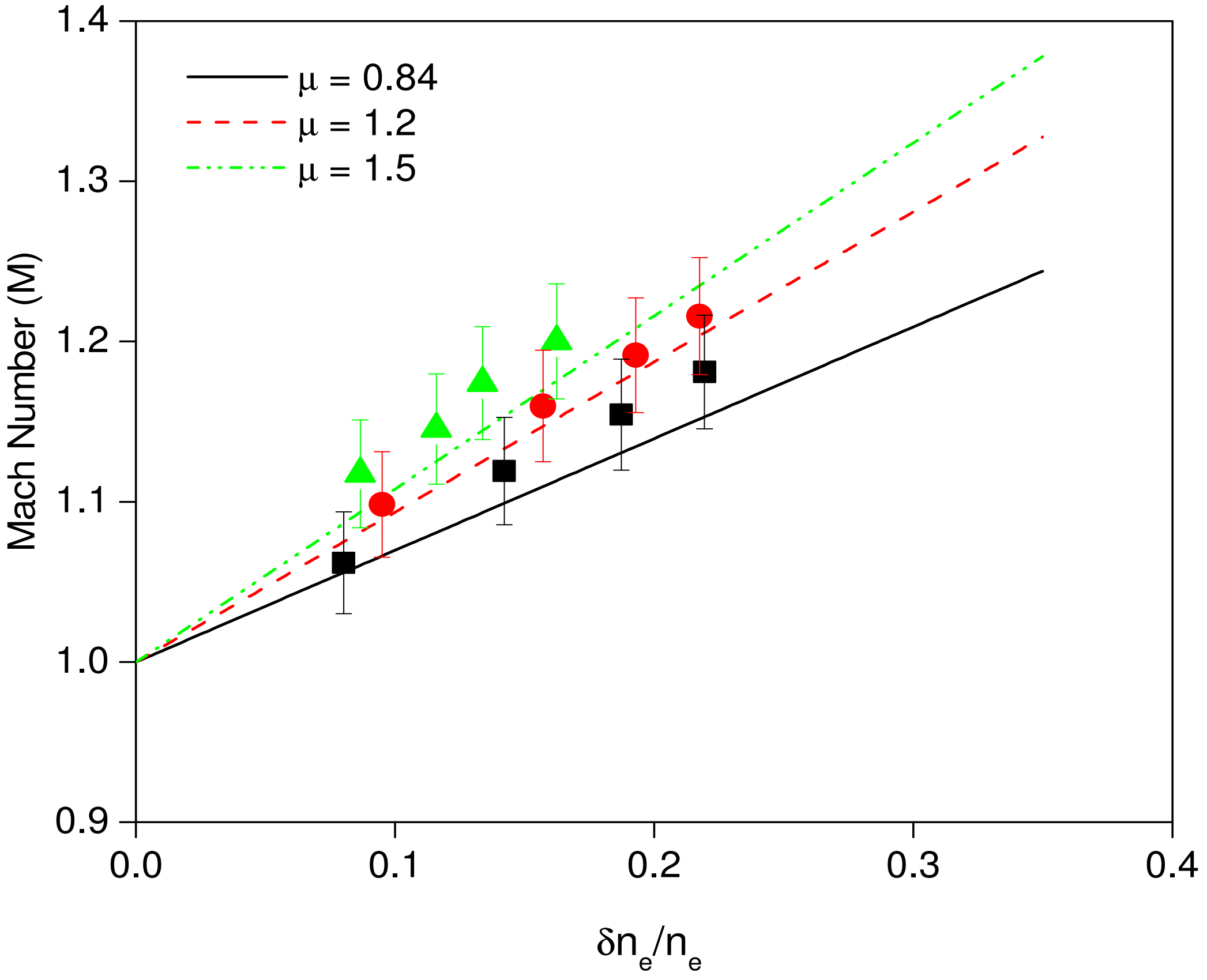

FIG. 8 (a). (Color on-line) Measured Mach number versus normalized amplitude of the compressive pulses with $\mu$ as a parameter (■ – $\mu$ = 0.84, ●–– 1.2, ▲–·· 1.5) in absence of dust.

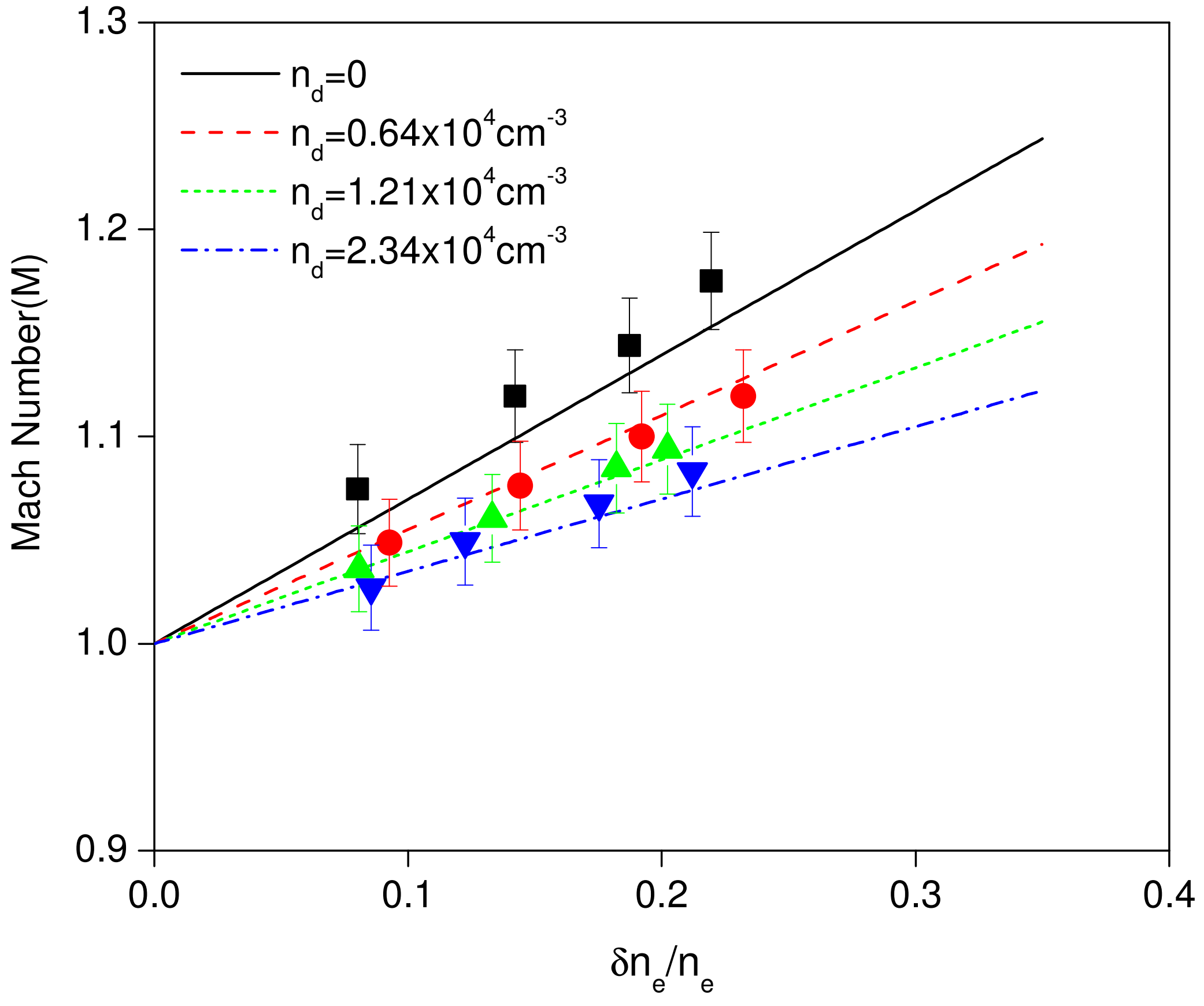

FIG. 8(b). (Color on-line) Measured Mach number versus normalized amplitude of the compressive pulses for an initial fixed value of $\mu( = 0.84)$ with dust density as a parameter (■ – $n_d = 0$ , ●––0.64 ×$10^4$ cm$^{-3}$, ▲--- 1.21 ×$10^4$ cm$^{-3}$, ▼–·–· 2.34 ×$10^4$ cm$^{-3}$).

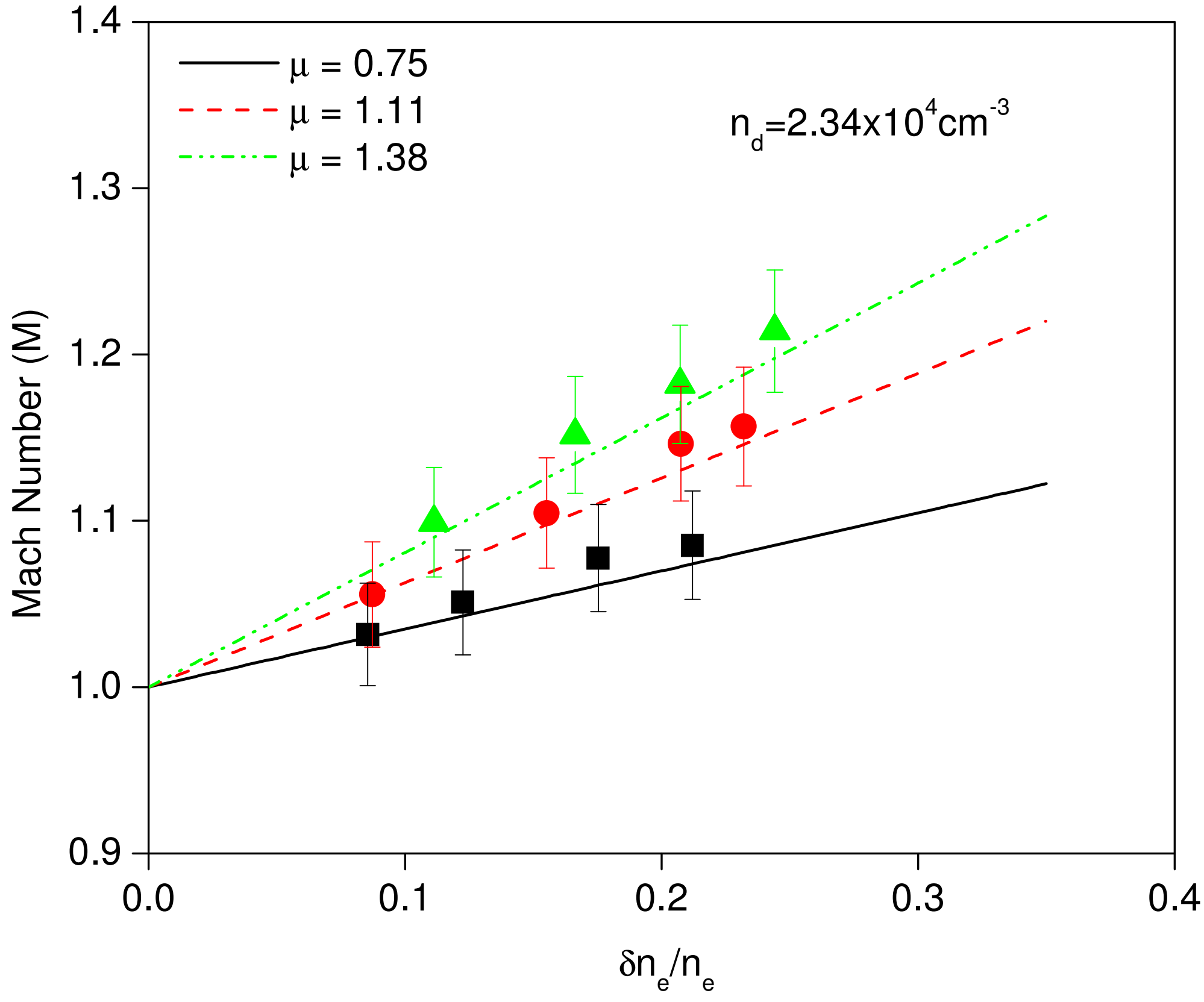

FIG. 8(c). (Color on-line) Measured Mach number versus normalized amplitude of the compressive pulses with $\mu$ as a parameter (■–$\mu$ = 0.75 , ●–– 1.11 , ▲–·· 1.38) in presence of dust.

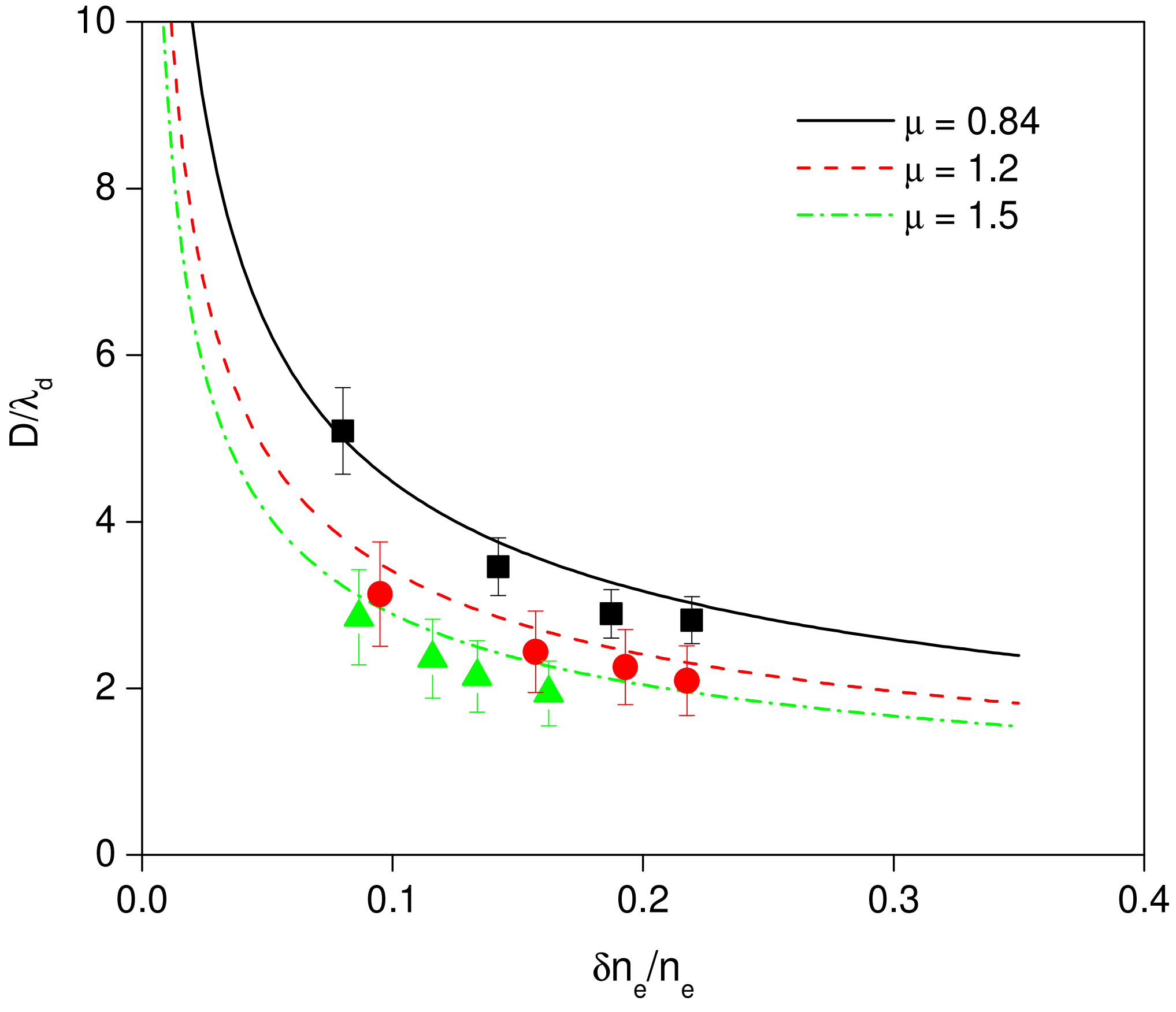

FIG. 9 (a). (Color on-line) Measured width versus normalized amplitude of the compressive pulses with $\mu$ as a parameter (■ – $\mu$ = 0.84, ●–– 1.2, ▲–·–· 1.5) in absence of dust.

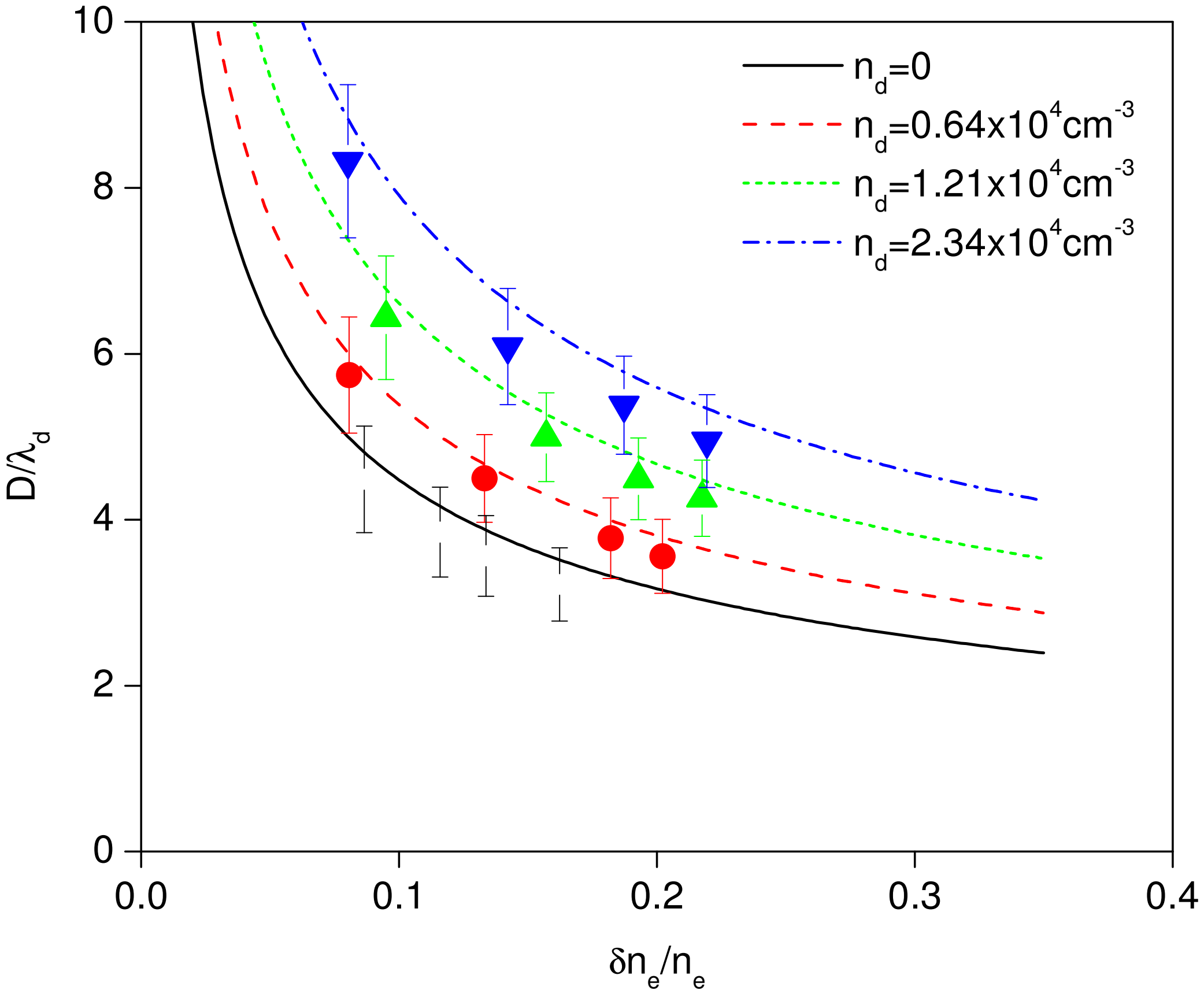

Fig. 9 (b). (Color on-line) Measured width versus normalized amplitude of the compressive pulses for an initial fixed value of $\mu$ (= 0.84) with dust density as a parameter (■ – $n_d = 0$ , ●––0.64 ×10$^4$ cm$^{-3}$, ▲--- 1.21 ×10$^4$ cm$^{-3}$, ▼–·– 2.34 ×10$^4$ cm$^{-3}$).

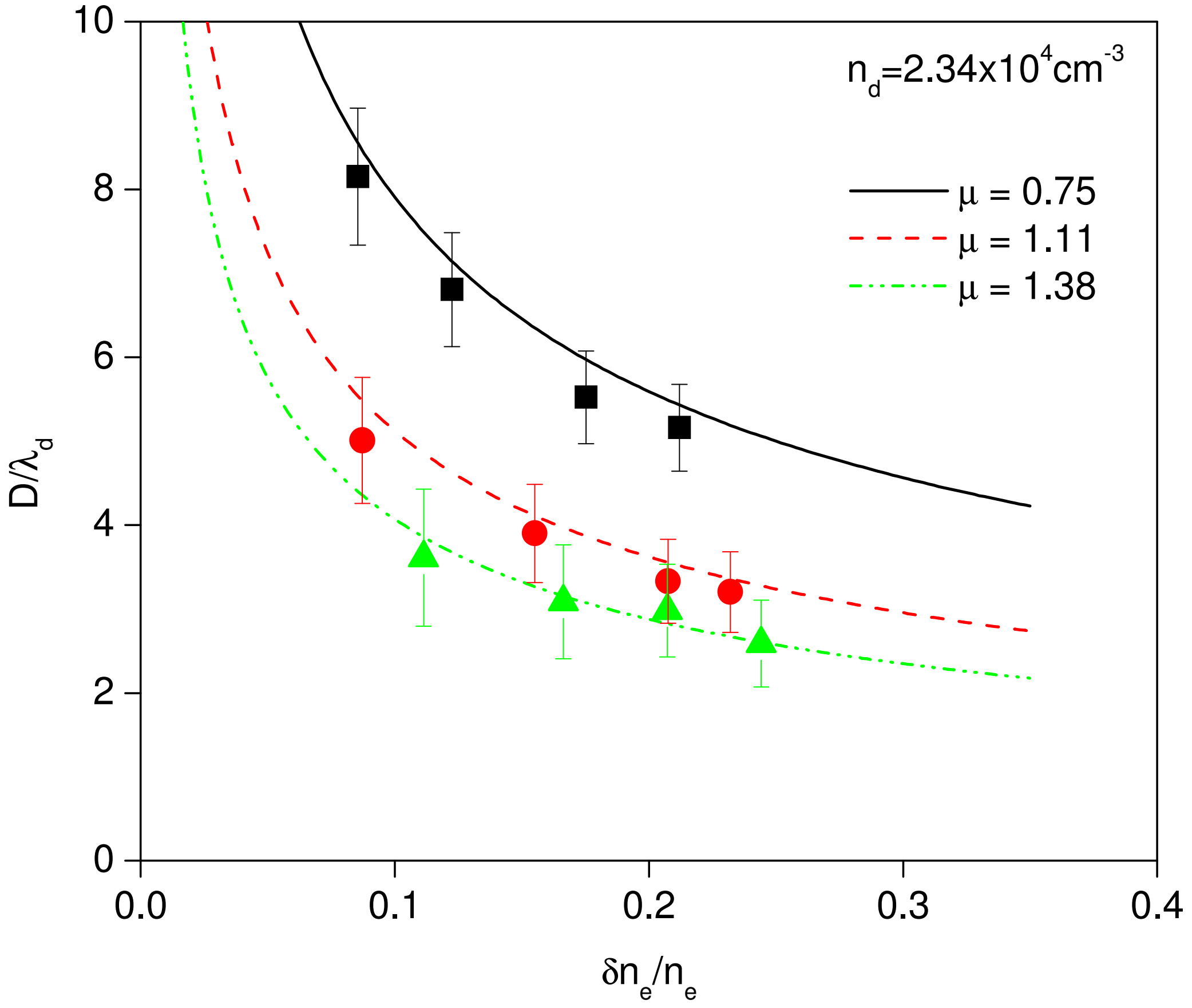

Fig. 9 (c). (Color on-line) Measured width versus normalized amplitude of the compressive pulses with $\mu$ as a parameter (■–$\mu$ = 0.75 , ●–– 1.11 , ▲–·· 1.38) in presence of dust.